\documentclass[prl,aps,epsfig,showpacs,twocolumn]{revtex4}
\usepackage{graphicx}
\usepackage{amssymb}
\usepackage{amsmath}

\begin{document}

\draft

\title{Rotation of easy axis in training effect and recovery of exchange bias in
ferromagnet/antiferromagnet bilayers }

\author{X. P. Qiu, D. Z. Yang, and S. M. Zhou}
\affiliation{Applied Surface Physics Laboratory (State Key
Laboratory) and Department~of~Physics, Fudan University, Shanghai
200433, China}

\author{R. Chantrell, K O'Grady, and U Nowak}
\affiliation{Physics Department, The University of York,York, YO10
5 DD, UK}

\author{J. Du and X. J. Bai}
\affiliation{National Laboratory of Solid State Microstructures,
Nanjing University, Nanjing 210093, China}

\date{\today}

\begin{abstract}

\indent  For ferromagnet/antiferromagnet bilayers, rotation of the
easy axis has been \textit{for the first time} observed during
measurements of training effect and the recovery of exchange bias
using FeNi/FeMn system. These salient phenomena strongly suggest
irreversible motion of antiferromagnet spins during subsequent
measurements of hysteresis loops. It is found that the rotation of
the easy axis can partly account for the training effect and the
recovery of the exchange bias.
\end{abstract}

\vspace{5 cm}


\maketitle

\indent After exchange bias (EB) is established in ferromagnet
(FM) /antiferromagnet (AFM) bilayers, hysteresis loop is shifted
along the horizontal magnetic field axis by an amount of exchange
field $H_{\mathrm{E}}$~\cite{1,2,3,4}. Meanwhile, the coercivity
$H_{\mathrm{C}}$ is enhanced, compared with that of corresponding
FM film. This phenomenon and other related physical properties
have been studied extensively, including rotational hysteresis
loss, training effect, asymmetrical magnetization
reversal, and rotational hysteresis of angular dependence of EB~\cite{3,9,Gao2007}.\\
\indent In general, $H_{\mathrm{E}}$ and $H_{\mathrm{C}}$ shrink
during subsequent measurements~\cite{3}. Although various
theoretical models have been proposed to explain the training
effect~\cite{Hoffmann2004,Neel1949,fulcomer,Mauri1987,Suess2003,Radu2003,Binek2004,Hauet2006},
the mechanism is unsolved. For example, in an early
approach~\cite{fulcomer,Zhang2001}, the training effect is
explained in terms of transition of spin orientation in AFM
grains. Based on Mauri model, however, the training effect is
ascribed to the irreversible motion of planar domain
wall~\cite{Suess2003,Radu2003}. Therefore, new experiments are
required to establish a unique
model of the training effect. \\
\indent Up to date, hysteresis loops are always measured
\textit{at the easy axis }(EA) in studies of the training
effect~\cite{Zhang2001,Brems2005}. At the EA, however, the
magnetization reversal process in the FM layer is often
accompanied only by motion of domain wall~\cite{9,Beckmann2003}.
In order to further reveal the nature of the training effect,
hysteresis loops should also be measured along other orientations,
at which the magnetization rotation occurs during magnetization
reversal process, in addition to the motion of domain wall. More
seriously, few other physical quantities have been measured in
studies of training effect. Actually, since the orientation of AFM
spins is altered during subsequent measurements of hysteresis
loops~\cite{Binek2004}, the orientation of the pinning field from
the AFM layer, i.e., the EA of the FM layer is expected to rotate.
In this Letter, we have \emph{for the first time} observed the EA
rotation, companied by the training effect and the recovery of the
EB in FM/AFM bilayers using FeNi/FeMn system. More remarkably, the
EA rotation can in turn account for the training effect and the recovery of the EB. \\
\indent A 1~cm $\times$ 5~cm bilayer of
Fe$_{20}$Ni$_{80}$(=FeNi)(3 nm)/Fe$_{50}$Mn$_{50}$(=FeMn) was
sputtered on glass substrate at ambient temperature. With a wedged
shape across the distance of 5 cm, the FeMn layer thickness
$t_{\mathrm{AFM}}$ is a linear function of the sampling location.
A uniform bilayer of FeNi(3 nm)/FeMn (2.4 nm) was also prepared. A
buffer layer of 15 nm Cu was used to stimulate the FCC (111)
preferred orientation in FeMn layers and to enhance
EB~\cite{Nakatani1994}. Finally, another 20 nm thick Cu layer was
used to prevent oxidation. The EB was established by a magnetic
field applied in the film plane during deposition. A
detailed fabrication procedure was given elsewhere~\cite{Gao2007}.\\
\indent X-ray diffraction shows that constituent layers are
polycrystalline with a strong FCC (111) peak and a weak FCC (200)
one. Before magnetic measurements, the large specimen was cut into
small pieces along the wedge direction. With a vector vibrating
sample magnetometer (VVSM), $m_{\mathrm{x}}$ and $m_{\mathrm{y}}$
were measured, as components of magnetic moment parallel and
perpendicular to the external magnetic field $H_{\mathrm{a}}$,
respectively. The $m_{\mathrm{x}}$ corresponds to conventional
hysteresis loops. $H_{\mathrm{a}}$, $m_{\mathrm{x}}$, and
$m_{\mathrm{y}}$ are in the film plane. As torque curves,
$m_{\mathrm{y}}$ were measured as a function of the
$H_{\mathrm{a}}$ orientation~\cite{Benito2006}. Under
$H_{\mathrm{a}}=0$, the EA can be identified as the angular
position of $m_{\mathrm{y}}=0$. All measurements were performed at
room temperature.
\begin{figure}[tb]
\begin{center} \resizebox*{7.2cm}{!}{\includegraphics*{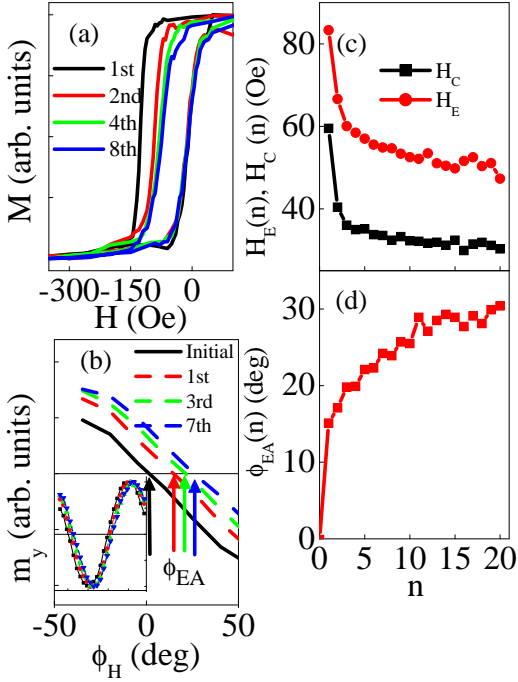}}
\caption[]{Hysteresis loops (a) and angular dependence of
$m_{\mathrm{y}}$ under $H=0$ with a small angular region (b),
dependence of $H_{\mathrm{E}}$ and $H_{\mathrm{C}}$ (c) and of
$\phi_{\mathrm{EA}}$ (d) on $n$  for uniform FeNi(3 nm)/FeMn (2.4
nm) bilayers, where $\theta_{\mathrm{H0}}=-8$ degrees for
measurements of hysteresis loops. In (b), arrows refer to
$\phi_{\mathrm{EA}}$. The inset shows the entire curve of
$m_{\mathrm{y}}$ versus $\theta_{\mathrm{H}}$ under $H=0$. }
\label{Fig1}
\end{center}
\end{figure}

\begin{figure}[tb]
\begin{center} \resizebox*{7.2cm}{!}{\includegraphics*{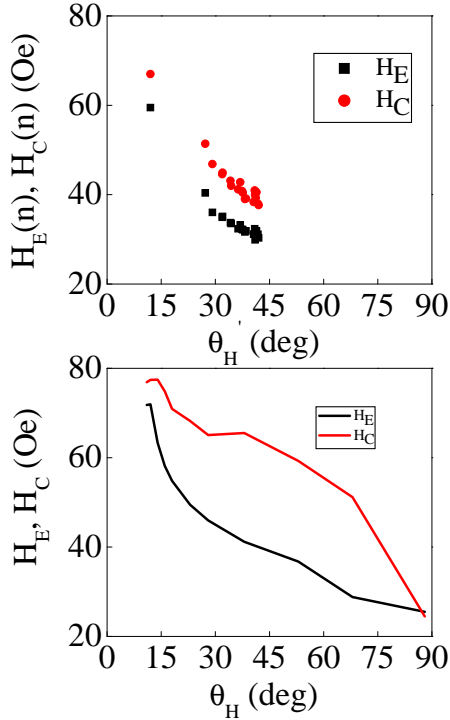}}
\caption[]{For uniform FeNi(3 nm)/FeMn (2.4 nm) bilayer,
$H_{\mathrm{C}}(n)$ and $H_{\mathrm{E}}(n)$ versus
$\theta_{\mathrm{H}}^{'}$ (a), and the dependence of
$H_{\mathrm{C}}$ and $H_{\mathrm{E}}$ on $\theta_{\mathrm{H}}$
(b). In (a) $\theta_{\mathrm{H}}^{'}=
\phi_{\mathrm{EA}}(n-1)-\theta_{\mathrm{H0}}$ and
$\theta_{\mathrm{H0}}=-8$ degrees.} \label{Fig2}
\end{center}
\end{figure}

\begin{figure}[tb]
\begin{center} \resizebox*{7.2 cm}{!}{\includegraphics*{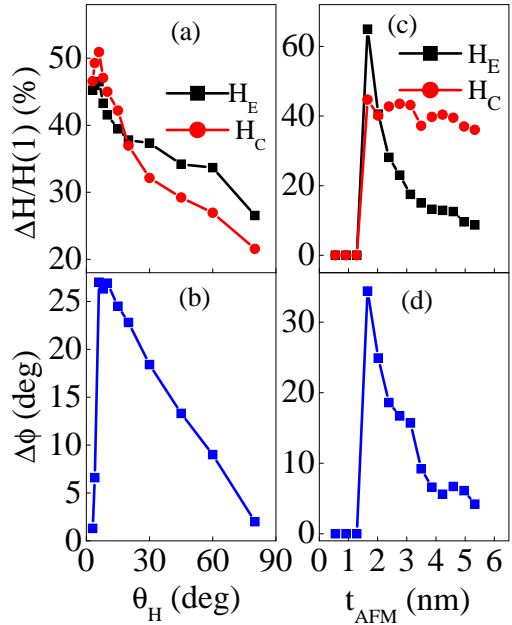}}
\caption[]{Dependence of $\Delta
H_{\mathrm{E}}/H_{\mathrm{E}}(n=1)$ and $\Delta
H_{\mathrm{C}}/H_{\mathrm{C}}(n=1)$ (a, c) and
$\Delta\phi_{\mathrm{EA}}$ (b, d) on $\theta_{\mathrm{H}}$ for
uniform FeNi(3 nm)/FeMn (2.4 nm) bilayer (a, b) and
$t_{\mathrm{AFM}}$ at $\theta_{\mathrm{H0}}$=-12 degrees for
FeNi(3 nm)/wedged-FeMn (0-6.25 nm) bilayers (c, d).} \label{Fig3}
\end{center}
\end{figure}

\indent For convenience, we use $\theta_{\mathrm{H}}$ and
$\phi_{\mathrm{EA}}$ to express the orientations of
$H_{\mathrm{a}}$ and the EA with respective to the EA at the
initial as-prepared state, respectively. For FeNi(3 nm)/FeMn(2.4
nm) bilayer, the EA at the initial state is at first identified.
Then, hysteresis loops at a specific $\theta_{\mathrm{H0}}$ and
torque curves with $\theta_{\mathrm{H}}$ from 0 to 360 degrees
under $H=0$ were alternatively measured. Figure~\ref{Fig1}(a)
shows hysteresis loops with subsequent measurements at
$\theta_{\mathrm{H0}}=-8$ degrees. The coercivity of decent branch
decreases significantly while that of ascent branch changes little
with the cycle number $n$. $H_{\mathrm{E}}$ and $H_{\mathrm{C}}$
decrease with increasing $n$, as shown in Fig.~\ref{Fig1}(c).
Although two subsequent hysteresis loops are interrupted by 3
minutes of torque measurements, $H_{\mathrm{E}}$ and
$H_{\mathrm{C}}$ change in a scale of $1/\sqrt{n}$, except for
$n=1$~\cite{3}. Moreover, in experiments, $m_{\mathrm{y}}$ at the
coercivity is found to increase with increasing $n$ (not shown).
For clarification, the torque curve, i.e., the angular dependence
of $m_{\mathrm{y}}$ in a small region is shown in
Fig.~\ref{Fig1}(b). Apparently, the position of
$m_{\mathrm{y}}=0$, i.e., $\phi_{\mathrm{EA}}$ is shifted towards
high angles. This also agrees with the results that
$m_{\mathrm{y}}$ at the coercivity increases with increasing
$n$~\cite{Spenato2007}. It should be noted that for the present
FeNi/FeMn bilayers, the uniaxial anisotropy axis and
unidirectional one are both aligned along the EA. This is because
$m_{\mathrm{y}}$ is always equal to zero for hysteresis loops
along the EA~\cite{Pokhil2004}. Figure~\ref{Fig1}(d) shows that
$\phi_{\mathrm{EA}}$ at first increases sharply with increasing
$n$ and then approaches saturation. Corresponding to much larger
$H_{\mathrm{E}}(n=1)$ and $H_{\mathrm{C}}(n=1)$,
$\phi_{\mathrm{EA}}(n=0)$ is much lower than those with $n \geq
1$, as predicted by Hoffmann~\cite{Hoffmann2004}. It is \emph{the
first time }that the EA in FM/AFM bilayers has been observed to
rotate
during subsequent measurements of hysteresis loops. \\
\indent To further reveal the nature of correlation between the
shrink of EB and the deviation of $\phi_{\mathrm{EA}}$, the
results in Figs.~\ref{Fig1}(c)~$\&$~\ref{Fig1}(d) are reorganized
showing the dependence of $H_{\mathrm{E}}(n)$ and
$H_{\mathrm{C}}(n)$ on $\theta_{\mathrm{H}}^{'}$ as a function of
$n$, where $\theta_{\mathrm{H}}^{'}$ is the angle between
$H_{\mathrm{a}}$ and the EA for the cycle number $n$ and equals to
$\phi_{\mathrm{EA}}(n-1)-\theta_{\mathrm{H0}}$ with
$\theta_{\mathrm{H0}}=-8$ degrees. As shown in Fig.~\ref{Fig2}(a),
$H_{\mathrm{E}}$ and $H_{\mathrm{C}}$ decrease with increasing
$\theta_{\mathrm{H}}^{'}$. We also measured the conventional
angular dependence of $H_{\mathrm{E}}$ and $H_{\mathrm{C}}$ on the
orientation of $H_{\mathrm{a}}$ for FeNi(3 nm)/FeMn (2.4 nm)
bilayers. For comparison, the angular dependence is shown within
the region from 0 to 90 degrees, as shown in Fig.~\ref{Fig2}(b).
The variation of $H_{\mathrm{E}}(n) $ and $H_{\mathrm{C}}(n)$ with
$\theta_{\mathrm{H}}^{'}$ agrees with the angular dependence of
$H_{\mathrm{E}}$ and $H_{\mathrm{C}}$ merely qualitatively. Sine
the deviation of the EA cannot account for the training effect
very well, however, effect of the exchange coupling energy between
FM and AFM layers and uniaxial anisotropy energy should be
considered, changes of which with $n$ were
observed in our experiments~\cite{Qiu008}. \\
\indent Here, we define $\Delta
H_{\mathrm{E/C}}/H_{\mathrm{E/C}}(n=1)$ to express the relative
change of $H_{\mathrm{E}}$ and $H_{\mathrm{C}}$ in training
effect, where $\Delta
H_{\mathrm{E/C}}=H_{\mathrm{E/C}}(n=1)-H_{\mathrm{E/C}}(n=20)$. In
a similar way, we have $\Delta\phi
=\phi_{\mathrm{EA}}(n=20)-\phi_{\mathrm{EA}}(n=0)$. Actually, for
$n>20$, the changes of $H_{\mathrm{E}}$, $H_{\mathrm{C}}$, and
$\phi_{\mathrm{EA}}$ are negligible, as shown in Fig.~\ref{Fig1}.
Figures~\ref{Fig3}(a)~$\&$~\ref{Fig3}(b) show the angular
dependence of the relative changes of $H_{\mathrm{E}}$ and
$H_{\mathrm{C}}$ and of $\Delta\phi$ for typical bilayer of FeNi(3
nm)/FeMn(2.4 nm). Firstly, as conventional
results~\cite{3,Zhang2001}, the training effect still exists at
$\theta_{\mathrm{H}}=0$ although $\Delta\phi=0$. Apparently, the
training effect is caused by the changes of the unidirectional and
uniaxial anisotropy energies instead of the EA rotation. Secondly,
at high $\theta_{\mathrm{H}}$, the training effect increases as
$\Delta\phi$ is increased as a function of $\theta_{\mathrm{H}}$.
Figures~\ref{Fig3}(c)~$\&$~\ref{Fig3}(d) show $\Delta
H_{\mathrm{E}}/H_{\mathrm{E}}(n=1)$, $\Delta
H_{\mathrm{C}}/H_{\mathrm{C}}(n=1)$, and $\Delta\phi$ as a
function of $t_{\mathrm{AFM}}$ at a specific
$\theta_{\mathrm{H0}}=-12$ degrees for FeNi (3 nm)/FeMn bilayers.
$\Delta H_{\mathrm{E}}/H_{\mathrm{E}}(n=1)$, $\Delta
H_{\mathrm{C}}/H_{\mathrm{C}}(n=1)$, and $\Delta\phi$ change in
similar variation trends. These scenario correlations indicate
that the training effect is mainly caused by the EA rotation in
the frame of the angular dependence of $H_{\mathrm{E}}$ and
$H_{\mathrm{C}}$ in Fig.~\ref{Fig2}(b).

 \indent For polycrystalline AFM layers, AFM grains are aligned randomly in the film plane.
For FeMn layers with (111) preferred orientation, AFM grains have
multi-easy axis anisotropy~\cite{Hoffmann2004,Urazhdin2005}. After
field-cooling procedure or at the as-prepared state under an
external magnetic field, AFM spins are expected to be aligned
along an EA close to the cooling field or the deposition magnetic
field. Assuming no interaction between AFM grains~\cite{fulcomer,
Gao2007}, AFM grains are suggested to have transitions from
non-equilibrium to equilibrium states triggered by subsequent
measurements of hysteresis loops~\cite{Binek2004,Brems2005}. As
the AFM spins of some grains have transition from one EA to
another one, the orientations of unidirectional and uniaxial
anisotropies are rotated~\cite{Olamit2007}. Meanwhile, their
magnitudes might also change. Therefore, $H_{\mathrm{C}}$ and
$H_{\mathrm{E}}$ at specific
$\theta_{\mathrm{H}}$ should shrink with $n$. \\
\indent In thin AFM layers, most of AFM grains are
"superparamagnetic" and the EB
disappears~\cite{Gao2007,Chantrell2000}. Hence, the training
effect and the deviation of the EA are equal to zero. At the
intermediate AFM layer thickness, most of AFM grains are thermally
stable, including rotatable and non-rotatable ones~\cite{11}.
Since a large fraction of AFM grains can be rotated irreversibly,
the deviation of the orientation of the effective pinning field
reaches maximum, so does the training effect. As
$t_{\mathrm{AFM}}$ is further increased, the volume of AFM grains
and the anisotropy energy barrier increase, resulting in a
reduction of transition possibility. The deviation of the EA and
the training effect are suppressed.  \\
\indent At $\theta_{\mathrm{H}}=0$, the magnetization reversal
process is accompanied only by the motion of domain
wall~\cite{9,Spenato2007}. AFM spins can only be switched by an
angular amount of 180 degrees~\cite{Beckmann2003} and the pinning
field is still aligned along that of the initial state and thus
$\phi_{\mathrm{EA}}=0$, as shown in Fig.~\ref{Fig3}(b). As
$H_{\mathrm{a}}$ is deviated away from the initial EA, the
magnetization reversal process is accompanied by both motion of
domain wall and magnetization rotation and finally by
magnetization coherent rotation for large
$\theta_{\mathrm{H}}$~\cite{9,Spenato2007}. Since AFM spins can be
rotated from one EA to other one, in addition to the 180-degree
switching, the EA can be deviated from that of the initial state.
With the FM magnetization rotation, the transition possibility,
thus the relative change of the EB and $\Delta\phi$ are
enhanced~\cite{Beckmann2003}, compared with those of 180-degree
switching. As $\theta_{\mathrm{H}}$ is further increased, the
magnetization reversal process for ascent and descent branches is
almost symmetric~\cite{9}. In this case, contributions of
transition possibility in the two pathways are cancelled so that
the deviation of the EA and thus the training effect are reduced.
In a word, $\Delta\phi$ strongly depends on $\theta_{\mathrm{H}}$
because the mechanism of the motion of AFM spins is determined by
the magnetization reversal mechanism of the FM layer. This latter
in turn depends on $\theta_{\mathrm{H}}$.

\begin{figure}[tb]
\begin{center} \resizebox*{7.2 cm}{!}{\includegraphics*{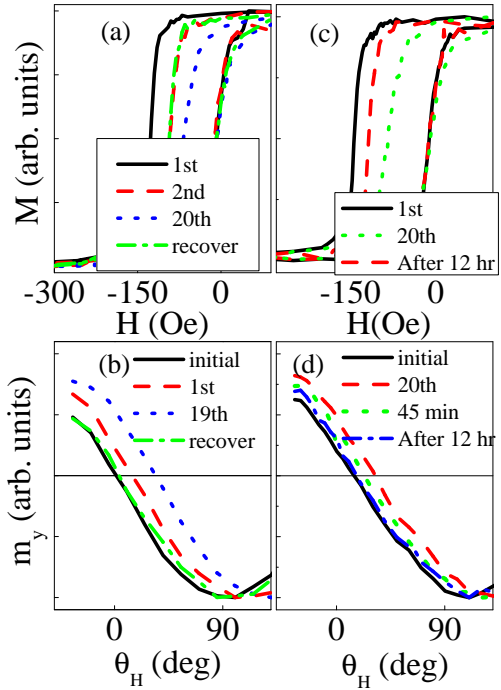}}
\caption[]{Hysteresis loops at $\theta_{\mathrm{H0}}$=-8 degrees
(a,c) and angular dependence of $m_{\mathrm{y}}$ under $H=0$(b,d)
using the first (a,b) and the second (c,d) recovery methods
 for uniform FeNi(3 nm)/FeMn (2.4 nm) bilayers.} \label{Fig4}
\end{center}
\end{figure}

\indent Although the EB recovery has been studied more
recently~\cite{Brems2005}, observation of the EA during the EB
recovery can elucidate the nature of this phenomenon. Here, the EB
recovery in FeNi(3 nm)/FeMn (2.4 nm) bilayers is studied using two
different methods. As the first method~\cite{Brems2005}, after
subsequent measurements of hysteresis loops at
$\theta_{\mathrm{H0}}=-8$ degrees, a hysteresis loop was measured
at an orientation perpendicular to $\theta_{\mathrm{H0}}=-8$.
After that, a hysteresis loop was measured again at
$\theta_{\mathrm{H0}}=-8$ degrees. As the second spontaneous
method, after $H_{\mathrm{C}}$ and $H_{\mathrm{E}}$ are stable
with subsequent measurements of hysteresis loops, $H_{\mathrm{a}}$
is set to zero for a designated period. Then, hysteresis loops and
torque curves were measured. In this way, $H_{\mathrm{C}}$,
$H_{\mathrm{E}}$, and $\phi_{\mathrm{EA}}$ are achieved at
different waiting time. Figures~\ref{Fig4}(a)~$\&$~\ref{Fig4}(c)
show that with either recovery method, $H_{\mathrm{C}}$ and
$H_{\mathrm{E}}$ are increased, compared with those of $n=20$.
Meanwhile, after the recovery procedure, the EA approaches back
towards that of the initial state, as shown in
Figs.~\ref{Fig4}(b)~$\&$~\ref{Fig4}(d). Therefore, the variation
of $\phi_{\mathrm{EA}}$, partly accounting for the EB recovery,
\emph{directly} verifies the micro-magnetic calculations that the
orientation of AFM spins are rearranged after recovery
procedure~\cite{Brems2005}. For CoO/Co bilayers, the orientation
of AFM spins is argued to change with respective to the cooling
field during training effect and recovery of the EB.  \\
\indent In summary, the EA in FM/AFM bilayers has been \emph{for
the first time} found to vary during the shrink and recovery of
the EB using FeNi/FeMn system. For AFM spins, the irreversible
motion of 180-degree switching or coherent rotation, depending on
the magnetization reversal mechanism of the FM layer, is
unambiguously demonstrated during measurements of hysteresis
loops. The training effect and $\Delta\phi$ vary in a similar way
with either $\theta_{\mathrm{H}}$ or $t_{\mathrm{AFM}}$.
Furthermore, the EA is rotated back towards that of the initial
state during the EB recovery. Therefore, the EA rotation is one of
the major reason for the training effect and the recovery of the
EB for large $\theta_{\mathrm{H}}$.\\
\indent Acknowledgement This work was supported by the National
Science Foundation of China Grant Nos. 50625102, 10574026, and
60490290, the National Basic Research Program of China
(2007CB925104) and 973-Project under grant no. 2006CB921300,
Shanghai Science
and Technology Committee Grant No. 06DJ14007. \\

\end{document}